# An Empirical Study on Secure Usage of Mobile Health Apps: The Attack Simulation Approach

Bakheet Aljedaani[a], Aakash Ahmad[b], Mansooreh Zahedi[c], M. Ali Babar[d,e]

[a]Computer Science Department, Aljumum University College, Umm Alqura University, Makkah, Saudi Arabia
[b]School of Computing and Communications, Lancaster University Leipzig, Germany
[c]School of Computing and Information Systems, University of Melbourne, Australia
[d]CREST – the Centre for Research on Engineering Software Technologies, University of Adelaide, Australia
[e]Cyber Security Cooperative Research Centre (CSCRC), Australia

[a]bhjedaani@uqu.edu.sa, [b]a.ahmad13@lancaster.ac.uk, [c]mansooreh.zahedi@unimelb.edu.au,
[d]ali.babar@adelaide.edu.au

**Abstract**

*Context:* Mobile applications (mobile apps for short) have proven their usefulness in enhancing service provisioning across a multitude of domains that range from smart healthcare, to mobile commerce, and areas of context-sensitive computing. In smart healthcare context, mobile health (mHealth) apps - representing a specific genre of mobile apps that manage health-critical information - face some critical challenges relating to security and privacy of device and user data. In recent years, a number of empirically grounded, survey-based studies have been conducted to investigate secure development and usage of mHealth apps. However, such studies rely on self-reported behaviors documented via interviews or survey questions that lack a practical, i.e. action-based approach to monitor and synthesise users' actions and behaviors in security critical scenarios.

*Objective and Method:* Our objective was to conduct an empirical study - engaging participants with attack simulation scenarios and analyse their actions - for investigating the security awareness of mHealth app users via action-based research. To conduct the study, we simulated some common security attack scenarios in mHealth context and engaged a total of 105 app users (14 countries across 5 continents) to monitor their actions and analyse their behavior. We analysed users' data with statistical analysis including reliability and correlations tests, descriptive analysis, and qualitative data analysis (i.e., thematic analysis method) for the open-ended questions.

*Results:* Our results indicate that whilst the minority of our participants perceived access permissions positively, the majority had negative views by indicating that such an app could violate or cost them to lose privacy. Users provide their consent, granting permissions, without a careful review of privacy policies that leads to undesired or malicious access to health critical data. The results also indicated that 73.3% of our participants had denied at least one access permission, and 36% of our participants preferred no authentication method.

*Conclusion*: The study complements existing research on secure usage of mHealth apps, simulates security threats to monitor users' actions, and provides empirically grounded guidelines for secure development and usage of mobile health systems.

## 1. Introduction

Mobile and pervasive technologies offer users with a multitude of context-aware services that include but are not limited to social networking, mobile commerce, along with smart and connected health care [25, 36]. Mobile health (mHealth) apps, and their enabling infrastructures such as context-sensitive mobile devices and wireless networking have empowered users and transformed the healthcare sector to provide a wide range of healthcare services. This has resulted in a significant improvement in the mHealth adoption by users (e.g., patients, medics, public health stakeholders), increased effectiveness, and reduced costs for healthcare services [25]. Public and private healthcare providers can leverage mHealth apps to offer digitize healthcare practices such as health and fitness monitoring [26], dermatologic care [17], chronic management [28, 59], and clinical practices [41]. A recent report by Allied Market Research revealed that the global digital health market size was valued at $145,884.3 million in 2020, and is projected to reach $767,718.9 million by 2030 [44].

Despite the potential benefits and strategic importance of mobile technologies and mHealth initiatives in smart systems context, a number issues such as resource poverty, security of device resources, and privacy of context-sensitive data represent critical challenges mobile computing solutions [24, 28, 36]. Specifically, the security of mHealth apps is considered a challenge due to the pervasive environment that continuously ingests health-critical data from embedded sensors (hardware), processes and persists data inside the device (via mobile apps), and transmits it across ad-hoc networks [36, 52]. According to a report by cybercrime magazine in 2020, ransomware has become the fastest growing and one of the most damaging types of cybercrime. The global cost for ransomware damage, including mobile devices, could reach $20 billion by 2021, which is 57 times more than it

was in 2015 [34]. As a typical example, some of the installed apps can be granted (undue/excessive) access unintentionally by the users to all of a device's resources to gain, use, and share users' data. Data in mobile devices and apps, specifically health-critical data in the context of this study, can be leaked to an external host or a third party through excessive app permissions [24, 62], phishing attacks [13], or when installing other mobile apps from unknown resources [11]. Recent studies (e.g., [6, 27]) highlight that users have limited security awareness of what they should do to protect their private and health-critical information. Besides, social engineering methods can be used by hackers to deceive users into leaking their private information [18, 62]. Molyneaux et al. in [32] indicated that it is difficult for users to make security-related decisions when facing security threats. Indeed, most users are unaware of such a threat to their data, or are unable to understand the technical mechanisms behind data leakage, which can lead them to ignore the associated security risks. Furthermore, employing suitable technical solutions including, privacy preserving mechanisms and two-factor-authentication, cannot address security issues alone; instead, the role of users and their understanding of how they should react to different security threats and attacks is an important factor in ensuring the secure usage of mobile apps [19].

In mobile systems - enhanced interactivity and context sensitivity are considered as ultimate criterion of success - technical measures alone may not be sufficient to ensure security, unless they are complemented with required human-centric knowledge and practices to protect data [3]. Human-centric knowledge and awareness in terms of necessary actions and behaviors, e.g., granting only the appropriate access permissions or opting-in for multi-factor authentication can significantly influence mHealth security [19]. According to a Proofpoint cybersecurity report, 95% of observed attacks exploited the "human factor" rather than relying on software and hardware vulnerabilities (such as phishing emails, or granting unnecessary permissions) [51]. Users could become a self-threat to their private data and can be easily deceived into revealing or leaking classified information if they are not fully aware of the security features they are utilizing [18, 62]. This issue can be seen with technology-based solutions that involve using security features in such a way that users find them difficult to understand [38]. mHealth app developers often undermine the user's security knowledge and behavior with a belief of having already delivered a secure app by following the principle and practices of secure software development [7]. However, users find these security features as hard to understand and use [18], confirmed with a recently conducted empirical study that investigated the security knowledge, attitude and behavior of mHealth app users [4]. The study highlights knowledge as the level of security understanding by the users, an attitude refers to how the users feel about their knowledge, and behavior refers to their actions that users perform to ensure security. The results of this study revealed that users had the knowledge, attitude about the security measures of the investigated apps. However, users' knowledge did not significantly influence their behaviors, indicating that users are aware of the risks but are reluctant, or unaware, of appropriate actions that mitigate such security risks (e.g., setting privacy preferences to restrict undesired data access).

To this end, state-of-the-art on secure usage of mHealth apps mainly relies on surveys and interviews to collect and analyze users' responses based on their security knowledge and behavior [4, 5, 21, 35, 57, 60]. However, the results of such studies could be affected by social desirability (i.e., self-reported behavior not supporting the respondents' claims), and/or the sampling method (i.e., relying on a particular group of users). The above-mentioned limitations demand for an action-based mechanism, taking users in the loop to monitor and analyze their actions in a real context, simulated as security threats scenarios to their health critical data. This study aims to investigate the behavior(s) of users (i.e., users' actions in a specific security scenario based on their knowledge) when dealing with mHealth apps. Users' actions and behaviors such as clicking on suspicious links or granting unnecessary permissions could lead to compromised security of data. Contrary to asking the users via interview(s) or survey(s) to report their behaviors that surfer measurement bias [43, 53, 54], we engage users to demonstrate their behavior that can be monitored and analyzed using an attack simulation approach. We aim find answers for two research questions that aim to investigate (1) the actions of users, and (2) mistakes made by them while using mHealth apps in a security critical scenario.

To find the answers, we adopted an action-based research approach [11, 15] to develop an attack simulation solution and engaged 105 (Android device) users to monitor their actions for analyzing their security knowledge. mHealth apps can be classified into different genres such as general health advisory to fitness monitoring and nutritional data apps, our study uses fitness monitoring app as such class of apps are considered as most common and widely used apps in mHealth systems. This study complements our previous works [4, 3] that aimed to understanding the security awareness of users while using clinical mHealth apps. This study involves participants from 14 countries (across 5 continents), with different age groups, various educational backgrounds, and different IT knowledge. Our decision to select the Android platform for security attack simulation solution is influenced by two reasons (a) it is the most widely used mobile platform for mHealth apps, and (b) it is considered as comparatively more vulnerable than other mobile platforms such as iOS. Study participants were monitored based on their actions and behaviors under simulated attacks on security of mHealth app. In addition, we sought users' input via an exit survey collecting user's demographic and other auxiliary information that compliments study results. Data analysis consisted of two steps. The first step involved statistical analysis to measure the reliability and correlations of study items, and a descriptive analysis to report the demographic data and our participants' (re-) actions corresponding to security threats. The second step included qualitative analysis to present the findings for the open-ended question. This study makes the following contributions:

- Simulates the most frequently encountered security threats to monitor and investigate the actions and behaviors of mHealth apps users – at attempt to overcoming the issues and bias in self-reported behaviors.
- Identify and classify the main reasons considered by users while facing access permissions' requests in mHealth context.
- Empirically grounded guidelines for secure usage of mHealth app driven by user's actions in security threat scenarios.

Based on the outlined contributions above, the study can have the following implications for researcher and developers of mHealth apps. More specifically,

- Researchers who are interesting in exploring security awareness of users, formulating new hypothesis to be tested for secure mHealth apps, and conducting action-based research on secure mobile computing.
- Developers and stakeholders to understand security specific knowledge and behaviors of user, monitored by our proof of the concept attack simulator and documented guidelines, to develop and provide emerging and futuristic class of mHealth apps that are secure and usable.

Section 2 presents the related work. Section 3 details the adopted research methodology. Sections 4 and 5 reports the study findings. Section 6 discusses the findings of the study. Section 7 describes validity threats for this study. Section 8 concludes the paper.

## 2. Related Work

We now review the most relevant existing research generally classified as (i) self-reported survey studies on secure mHealth apps (Section 2.1) and (ii) action-based research to understand secure usage of mobile computing (Section 2.2), detailed below. The review follows a conclusive summary to contextualize the scope and contributions of the proposed research.

### 2.1 Self-Reported Surveys to Investigate Security Awareness of mHealth Users

In the extent literature on mHealth security, one of the most common approaches to assess the security awareness of users is to conduct survey-based studies (i.e., survey, interviews, focus groups, etc.). Such approaches for survey-based studies [4, 5, 21, 35, 57, 60], engage mHealth app users to self-report their security-related actions and behaviors in the past and the consequences of such actions [56] via pre-defined questionnaire. From an empiricism perspective, such self-reported studies may suffer some inherent bias including but not limited to exaggerated answers, or a sense of scepticism (often shyness or embarrassment) in sharing private information that may relates to someone health critical data [31], resulting in unreliable findings [11].

*Security actions and habits:* A recently published work [4] surveyed 101 participants to measure the security awareness of users about using clinical mHealth apps. The study utilized the Human Aspects of Information Security (HAIS) model to measure users' security knowledge, attitude, and behavior. Besides the potential bias in data collection (i.e., self-reported behavior), the study participants were from only two health providers in one region that impacts participants diversity, and ultimately impacting the generalisation of the results. The study by Zeybek et al. [60] surveyed 120 participants in a public institution to investigate mobile apps users' security habits (i.e., installation and rejection for app's permissions, the usage of antivirus app, locking screen, frequent updates for the apps, and installation for the apps from official store). The study concluded that awareness training and malware analysis through internal experts or external institutions are required. The results can be biased due to the fact that employees might report their best behavior as they have been informed that the study outcomes might be view by their senior managers.

*Cyber Security awareness:* Watson et al. [57] surveyed 94 participants and Mylonas et al. [35] surveyed 458 participants to investigate the security awareness of mobile devices users about critical security options (e.g., device settings, user behaviors, etc.). While the findings of Watson et al. revealed that participants, especially those without strong IT knowledge, tend to ignore or are unaware of many critical security options, Mylonas et al. found that users were not adequately prepared to make appropriate security decisions. Furthermore, users had poor adoption of 'pre-installed' security controls, such as encryption, remote data wipe, and remote device locator. The limitations that could affect the results can be relying solely on the self-reported behavior, and focusing on specific group of participants (age 15 – 30). A study by Alotaibi et al. in 2016 [5] surveyed 629 Saudi Arabia based participants to investigate the cyber security awareness of computer and mobile users by focusing on three contexts, namely, cyber security practices, cybercrime awareness, and incident reporting. The study found that, although the participants had a good knowledge of IT, their awareness of the threats associated with cybercrime, cyber security practices, and the role of government and organizations in ensuring information safety across the internet, is very limited.

## 2.2 Action-based and Experimental Research on Secure Mobile Usage

The second approach is measuring the security awareness using action-based research (other studies call it objective data sources) such as [8, 11, 15, 16, 47, 58]. This approach tends to measure the actual behavior through installing an agent and allows the researcher to monitor the participants' reactions of a specific security phenomenon. Producing inaccurate results is one of the limitations for survey-based research due to the self-reported behavior. Thus, studies developed more effective mechanism to measure the security awareness through measuring the actual behavior of users, namely simulation-based approach [11, 15].

*Action-based Understanding of App Access and Permission Actions:* In addition to action-based analysis, studies such as [16, 58] were found involving other methods, (e.g., interviews, think aloud, exit survey) to allow participants to share their thoughts instead of relying on simulation-based approach only. Specifically, the researchers in [58] conducted an experiment with 36 participants to examine users' ability to deny applications access to protected resources. Felt et al. [16] evaluated whether or not Android users pay attention to, understand, and act on permission information during installation. The study conducted through two usability studies: an Internet survey of 308 Android users, and a laboratory study wherein 25 participants interviewed and observed.

*Experimental Studies to Understand Secure Usage of Mobile Apps:* Bitton et al. [11] measured the security awareness of smartphone users (i.e., 162 participants) for specific attack classes. The actual behavior was measured using a developed mobile agent and network traffic monitor and compared the findings with self-reported behavior, which have been collected through a survey. Barth et al. [8] examined the privacy paradox by focusing on the actual behavior and eliminating the effects of a lack of technical knowledge, privacy awareness, and financial resources. The study conducted as an experiment (including a questionnaire) on the downloading and usage of a mobile phone app among 66 Computer science students by giving them sufficient money to buy a paid-for app. Egelman et al. [15] measured the computer and mobile device security attitudes of users by utilizing the Security Behavior Intentions Scale (SeBIS). Four security sub-scales were mainly investigated including awareness towards phishing attacks, passwords, frequent updates, and locking the devices. The study was conducted through two surveys (555 participants) and a field study to monitor the security activities of 71 participants. The study found that: (i) testing high on the awareness sub-scale correlated with correctly identifying a phishing website, (ii) testing high on the passwords sub-scale correlated with creating passwords that could not be quickly cracked, (iii) testing high on the updating sub-scale correlated with applying software updates, and (v) testing high on the securement sub-scale correlated with smartphone lock screen usage (e.g., PINs).

**Conclusive Summary:** We now present a conclusive summary and comparative analysis of the most relevant existing research, as in Table 1. Comparative analysis is based on four-point criteria including *(i) research challenge(s), (ii) focus and contributions, (iii) evaluation context,* and *(iv) research limitations*. The study reference points to an individual research work under discussion and its year of publication. Studies including [4, 5, 21, 35, 57, 60] measured the security awareness of users about using mHealth apps. However, all the indicated studies were measuring security behavior through a questionnaire, which is not sufficient to report accurate measurement. Alternatively, studies including [8, 11, 15, 16, 47, 58] tend to measure the security awareness using action-based research that can provide better results. However, none of the action-based research studies were focused on mHealth apps. To the best of our knowledge, there has been no experimental research study to measure users' security awareness about using mHealth apps. We investigated the security awareness for users of mHealth apps via an attack simulation approach. We measured users' reactions through posing a few security threats, and monitor their spontaneous reactions. Furthermore, our study followed by an exit survey to collect demographic data from the participants and capture their security views.

Table 1. Comparative Analysis of most Relevant Existing Studies Compared to our Study

| Study Reference | Research Challenges | Focus and Contributions | Evaluation Context | Research Limitations | Pub. Year |
|---|---|---|---|---|---|
| **Survey-based Studies** | | | | | |
| [4] | To exploit the Human Aspects of the Information Security (HAIS) model to investigate the security awareness of users regarding the usage of mHealth apps. | - Analyse users' *security knowledge, attitude, and behavior* towards mHealth apps<br>- Investigate prominent security issues for users such as privacy and usability. | - Survey of mHealth app users (*101 participants*)<br>Quantitative data | - Bias in data collection (diversity and types of participants)<br>- *Self-reported behavior* by users | 2020 |
| [60] | To investigate, through empirical study, the *security awareness of public institution personnel* towards using mobile devices. | - Investigate Security habits of users (*apps' permissions, usage of antivirus and protection mechanisms, lock screen, etc.*) | - Survey of workspace users (*120 participants*)<br>- Quantitative data | - *Bias in data collection*<br>- *Self-reported intentions of users* | 2019 |
| [57] | To analyse the usage of security settings and control by mobile device users. | - Analyse security recommendations to users (e.g., *device settings, user behaviors, and applications*) | - Survey of mobile app users (*94 participants*)<br>- Survey-based qualitative data.<br>- Quantitative data | - Bias in data collection (self-reported behavior of users)<br>- Diversity of users (i.e., students, faculty, and staff of one institute) | 2017 |
| [5] | To conduct an empirical Investigation into the cyber security awareness of users | - Analyse challenges of cyber security.<br>- Investigate cyber security awareness (*cyber security practices, cybercrime awareness*, and *incident reporting.*) | - User survey questionnaire (*629 users*)<br>- Quantitative and qualitative data | - Data collected from users of mobile and PCs.<br>- Bias in data collection (self-reported intention) | 2016 |
| **Controlled Experiments** | | | | | |
| [8] | To investigate the privacy and security paradox of mobile users by focusing on the actual behavior. | The study eliminated the effects (i.e., *lack of technical knowledge, privacy awareness, and financial resources*). | Experiment and survey (*66 participants*) | - Bias in recruiting users (High knowledge in IT).<br>- App selection might affect participants' consideration of security and privacy. | 2019 |
| [11] | To classify types of security threats and investigate information security awareness of mobile users for different classes of threats. | - Measuring security awareness of users<br>- Measuring users' behavior by mobile agent and network traffic monitor | - Survey of the mobile user (*162 participants*)<br>- Monitoring of mobile agents and mobile network traffic | - No specific type of apps to be investigated | 2020 |
| [15] | To exploit the Security Behavior Intentions Scale (SeBIS) to analyse the security attitudes of users | - Analyse the security attitude of users<br>- Supporting users' security knowledge (e.g., awareness *of phishing attacks, frequent updates, passwords*, and *device locking*). | - User survey questionnaire (*555 participants*)<br>Experimental monitoring of users (*71 participants*). | - No specific type of apps to be investigated<br>- Bias in data collection (creating passwords for low-risk accounts) | 2016 |
| **Proposed Study** | To monitor the security awareness of users of mHealth apps via a security attack simulation. | - Understanding users' security behavior when they are facing certain security threats/attacks. | - Utilising a simulation system and a survey to collect data. Quantitative and qualitative data | - Focusing on mHealth apps.<br>- Multiple data sources (i.e., simulation-based, survey)<br>Engaging users with diverse backgrounds (e.g., age, education level, etc.). | NA |

## 3. Research Method

In this section, we discuss the research methodology that we followed to investigate the security awareness of users of mHealth apps through a simulation-based attack. The adopted method comprises of three phases, each detailed below, as per the illustrations in Figure 1.

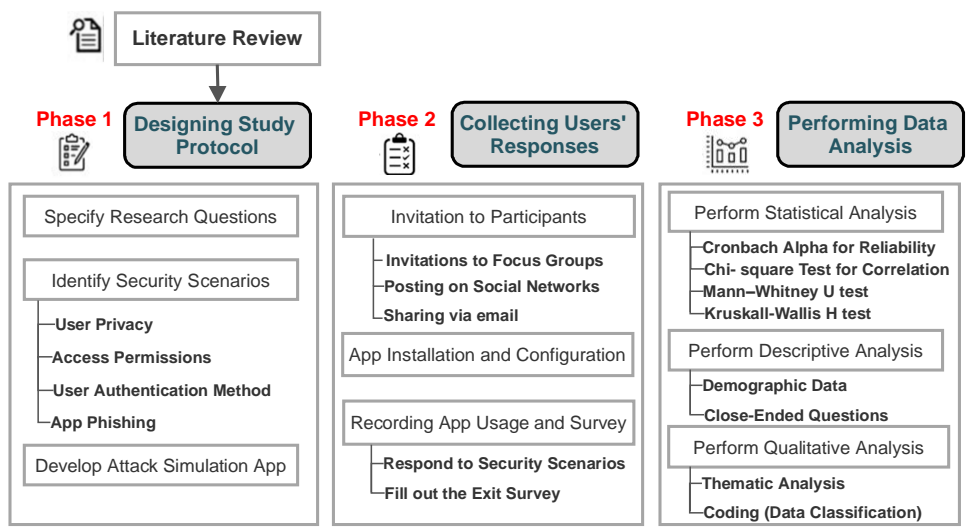

Figure 1. Overview of Research Method for Simulation-based Attack

### 3.1 Phase 1 – Designing the Study Protocol

Our literature review in Section 3.2 helped us understand state-of-the-art on users security issues, concerns, and preferences when using mHealth apps. In addition, we reviewed the literature to identify the existing methods to measure users' security awareness when using mobile apps (Section 3.3). Consequently, we understand and analyze the impact of the current research, highlighting proposed solutions and their limitations, specifying the research questions, and designing our study as in Figure 1. In the study protocol, we (i) specified the research questions (**RQ1**, and **RQ2** outlined in Section 1) (ii) identified the security scenarios that we plan to investigate, as in Table 2, and (iii) developed the simulation app. We investigated three major security threats, including user privacy and app permissions, authentication, and application phishing. It should be noted that each major security threat has a few potential security threats and at least one attack scenario, as highlighted in Table 2. We also gave the participants the option of reporting any security threats that they found and their interest in receiving security advice or training. Furthermore, we concluded our experiment with an open-ended question to capture users' thoughts in regards to the reasons behind paying attention to app permissions and the impact of giving an app more access permissions.

*Identification of Security Scenarios:* Studies such as [4, 3, 6, 39, 61] examined users' views about the security of mHealth apps. The surveyed participants agreed that mHealth apps need to implement the security countermeasures that ensure confidentiality, availability, and integrity of health-critical data. However, there is a lack of evidence based on what participants' reactions are in a real scenarios. Studies such as [8, 11, 15, 16, 47, 58] conducted experimental research to measure the security behaviors for users of mobile apps when they face certain security challenges (e.g., phishing attacks, app permissions). These studies have inspired our research and significantly helped us to identify four security threats and eight security scenarios to be examined as illustrated in Table 2. To further understand the participants' reaction about these security threats, we included an option to report the security issues that they may notice. In addition, we asked our participants in case they want to learn about security which can help us to plan further research. This study is approved by the Human Research Ethics Committee at the *University of Adelaide (H-2021-106)*. It should be noted that due to the conditional ethics approval, we omitted some security scenarios (e.g., requesting unnecessary information from participants, investigating password strength and updating password) to ensure participants' anonymity, avoid collecting sensitive data such as passwords.

Table 2. The Investigated Security Threats and Scenarios in the Simulation-based Attack Study

| Major security threats | Minor sub-security threats (Potential Security Threat) | Example (Scenarios/use-cases) |
|---|---|---|
| A. User Privacy and Permissions | 1. Reading privacy policy | • Show privacy policy to users to investigate the time that spent to read it. |
| | 2. Request access permission to device resources (e.g., user location, contacts, photos, microphone, camera, data of other apps) | • Request access permissions that are not mandatory for the app. The given permission for users can be either accept, or deny. |
| B. Authentication | 3. Selecting secure authentication methods (e.g., none, PIN, 2FA, etc.). | • Provide a few options for authentication to investigate users' preferences. |
| C. Application Phishing | 4. Pop-up window that requests users to share their private information. | • Show a pop-up window to request information from users given them the options to allow pop-up, or discard pop-up. |
| D. Feedback and reporting | 5. Reporting security issues to developers. | • Ask participants if they want to contact developers to report any security issues. |
| | 6. Understanding users interest in security education and training. | • Provide a signup option to receive frequent emails on secure mobile app usage. |

***Simulation Mobile App:*** We developed a health and fitness app called "Workout" to convince participants to be a part of the study. Due to time limit that we had to complete this research, we made the app (.APK file) downloadable directly for our storage instead of uploading it to apps stores. In fact, some participants were in doubt about installing the simulation app in their devices. The .APK file is available in [2]. Thus, we assured them that this is a part of a research study and we received ethical approval from concerned authorities and institutes to conduct this research and that the collected data will be anonymous and will be used solely for research purposes. We developed the simulation app to attract the participants through providing useful content such as workout plans, specific muscles exercise, etc. As an appreciation of our participants' time, we gave full access to the app (i.e., no subscription fees, no ads within the app) by removing all things that related to the study and make available on the app markets. A promotion code were given to all participants to use it for free.

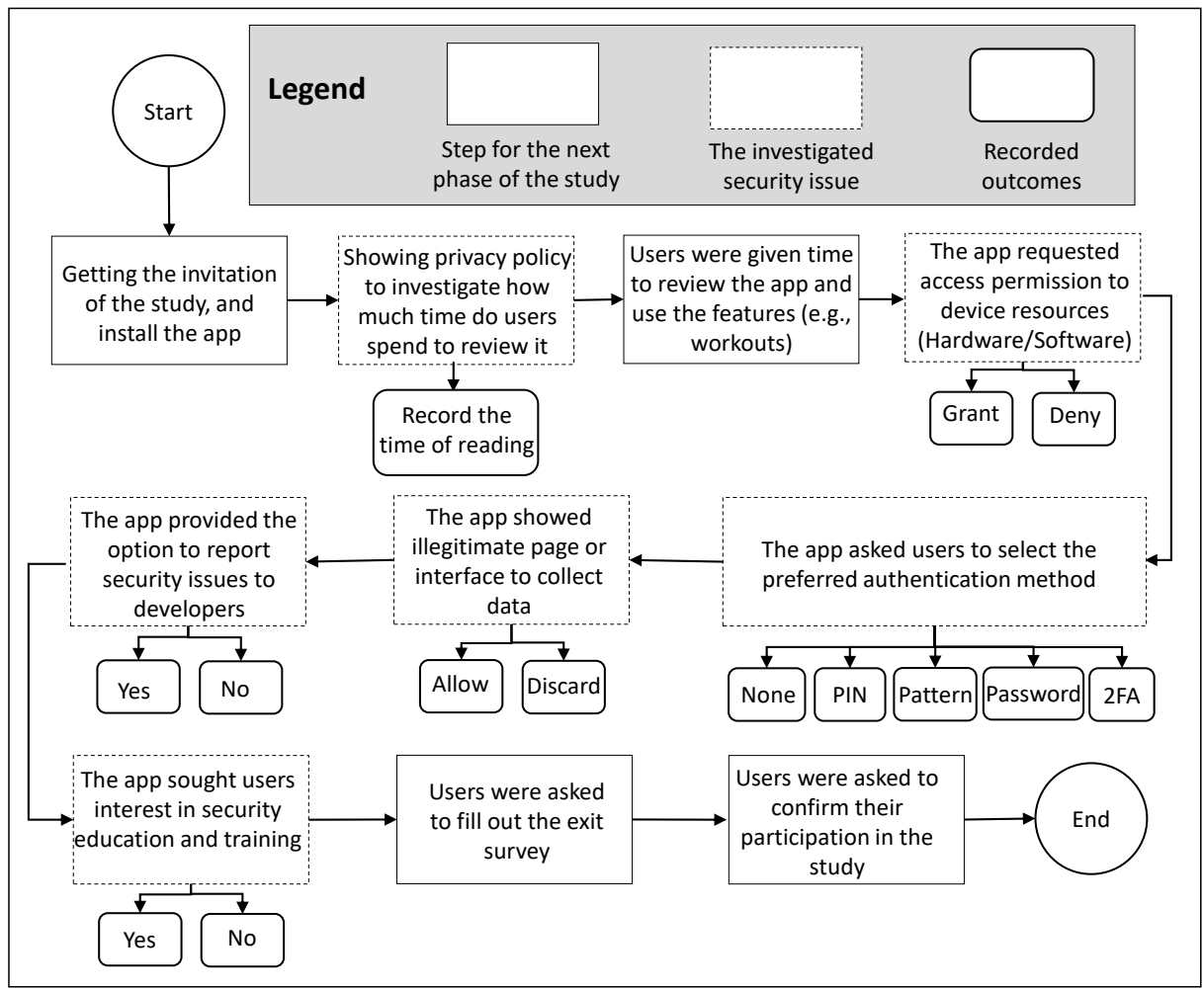

Figure 2. Flowchart of our Study Procedure

## 3.2 Phase 2 – Collect Users' Responses

***Recruitment for Participants:*** We targeted anyone who uses Android device since our simulation app is only compatible with Android OS. We advertised for our study by posting our study invitation on social network (e.g., Twitter, Facebook, LinkedIn, Reddit, WhatsApp groups, etc.) to recruit participants. Our app was available to download by visiting a designed webpage that also include instructions to involve in the study. Participant has to be at least 18 year-old to engage in the study. We were able to monitor participants' reactions from the moment they install the app and we recorded their security reactions. Participants faced many challenges related to the security aspects of mobile health apps, as illustrated in Table 2. Figure 3 shows six examples (screenshots) of the research activities that our participants faced. Additional screenshots for our simulation app are available in [2]. At the end, we provided an exit survey to collect demographic data and an open-ended question to allow them to share their thoughts. It should be noted that we carried out a pilot testing for our study with at least 10% of the participants to ensure the reliability and validity. Participants were given the option to withdraw their data during the study by simply deleting the app. As a result, their responses would be considered as incomplete and hence excluded from further analysis. After removing incomplete responses (i.e., we consider the response is complete when the participants went through all security scenarios which we are investigating, and filled out the exit survey), we were able to collect data from 105 participants (referred to as **P1** to **P105**). Collected data along with the exit survey questions are available in [2].

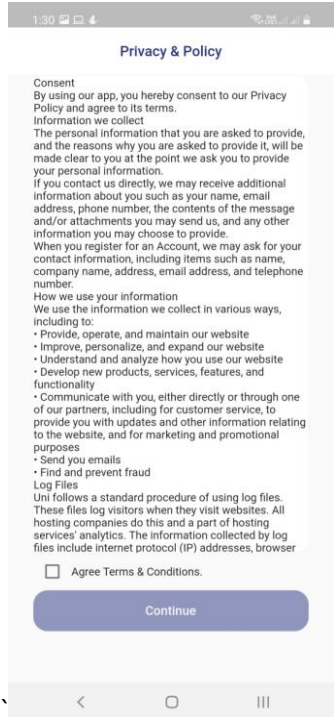
(a) Privacy Policy for our simulation app

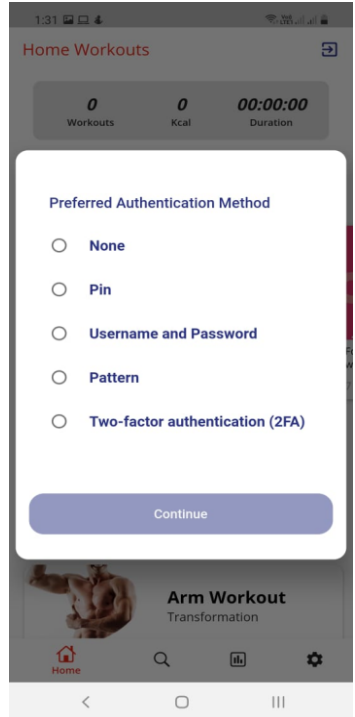
(b) Selecting the preferred authentication method

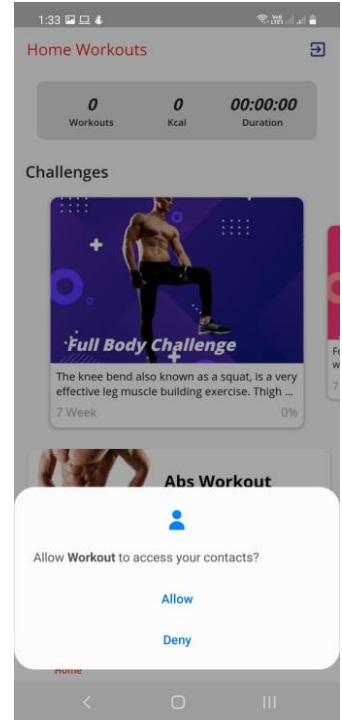
(c) Example of requesting access permission from participants

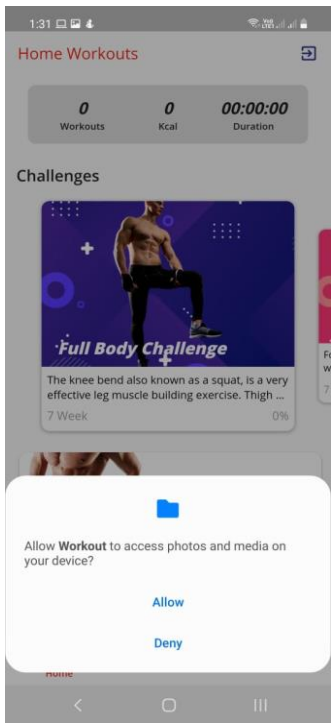
(d) Example of requesting access permission from participants

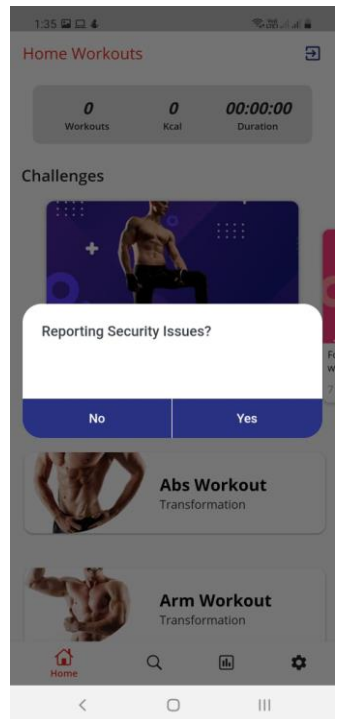
(e) Reporting security issue about the simulation app

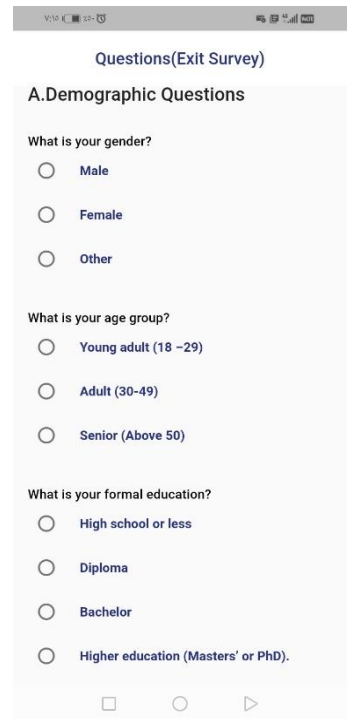
(e) Exit survey for the simulation app

Figure 3. Examples of Research Activities for Simulation-based Attack Study

***Participants' characteristics:*** As depicted in Figure 4, our study engaged participants from various countries (i.e., 14 countries, five continents), and with different age groups (participants should be at least 18 years old). Furthermore, our participants were having various educational backgrounds and different IT knowledge. The demography analysis presented in Figure 4 complements the exit survey responses (Q1-Q6), available in [2], helps us with fine-grained analysis of the study results. For example, the analysis helped us to understand the difference in male and female reactions about access permissions. We were able to recruit 105 Android devices users. Figure

4 (a) present the geographical distribution of our study participants. 64% were male, and 36% were female as in Figure 4 (b). As in Figure 4 (c), 45% were aged (18 – 29), 50% were aged (30 – 49), and only 5% of our population were above 50. The bachelor's degree is the most reported level of formal education for our respondents (i.e., 50%). 26% were postgraduate level, 13% were having diploma, and 10% were having high school or less, as in Figure 4 (d). In regard to the respondents' IT knowledge, 46% reported that they have little or no knowledge, 40% considered themselves having moderate knowledge, and only 14% had advanced knowledge, as in Figure 4 (e).

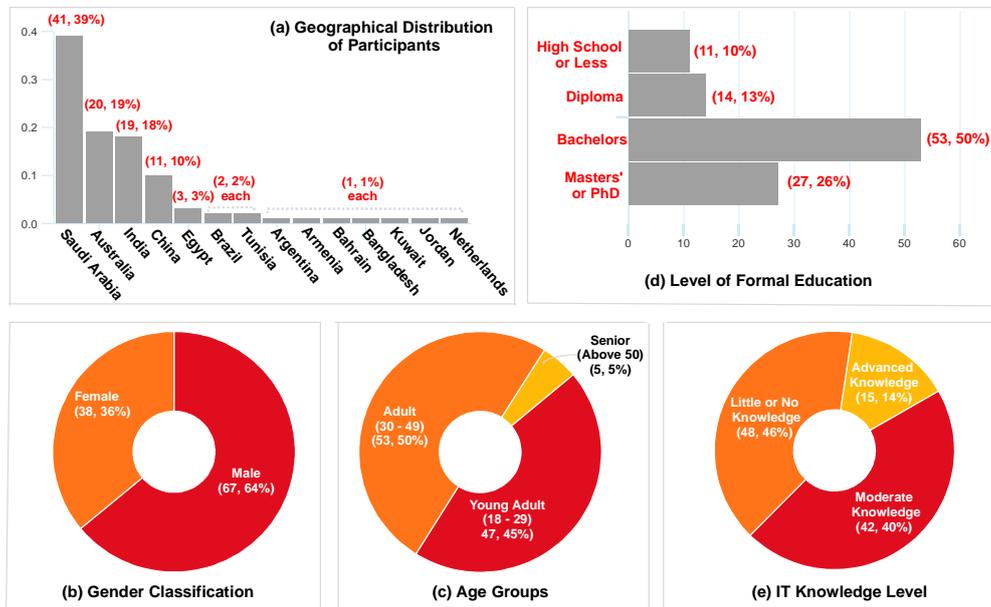

Figure 4. Demographic Details of the Participants of Attack-Simulation Study (Sample Size =105)

### 3.3 Phase 3 – Perform Data Analysis

For this study, we collected quantitative and qualitative data that help us to answer the outlined RQs. Based on the nature of data and the purpose of the analysis, we performed three techniques. Each technique is detailed below as per the illustrations in Figure 1.

*Statistical analysis:* SPSS version 28 (a popular data analysis software) was used in this technique. To measure the reliability and internal consistency of the dichotomous study items (i.e., permission requests), we calculated the **Cronbach Alpha** [49] to ensure that we meet the minimum acceptable coefficient value (i.e., ≥ 0.7). We also evaluated the correlation between the permission requests responses using **Chi-square test** [45]. Chi-square test helped us to understand the correlation (i.e., strength and direction of association) between the permission requests items. Furthermore, we used statistical significance to compare our participants' reactions to the requested access permissions among multiple groups of users. **Mann–Whitney U test** was performed to compare the gender group since we have two independent samples (i.e., male and female) [9]. The **Kruskall-Wallis H test** was performed to compare the participants' reactions for more than two independent samples (e.g., IT knowledge level, age group, etc.) followed by a post hoc test (once required) to examine the significance of differences in the mean scores for the specific group of users [22, 46]. For each demographic data, we tested the null hypothesis (i.e., *H0: there is no significant difference*) against the alternative hypothesis (i.e., *H1: there is a significant difference*), whereas $\mu_1, \mu_2, \ldots, \mu_k$ refers to population means.

*Descriptive analysis:* We reported the participants' demographic data by using **descriptive analysis** as presented in Figure 4. We also used this technique to report the participants' reactions when we asked them to make certain security decisions during using the simulation app (e.g., requesting user to select the preferred authentication method, requesting access to device storage, etc.).

*Qualitative analysis:* For the open-ended question, we used **thematic analysis** method [14] to identify the reasons that users reported about paying attention to the requested access permissions. Thematic analysis supports extracting the data and synthesizing the results. We used NVivo software, a popular computer-based tool, to organize and analyze data. It should be noted that coding was initially done by one of authors in the team that was reviewed and revised (wherever required) by the third author to avoid potential bias.

**Findings of the study:** We now present the findings to answer **RQ1**, and **RQ2** as outlined in Section 1. We divided our results section into two main sections: (i) Section 4 that reports the statistical and the descriptive analysis, and (ii) Section 5 that provides a thematic analysis for the open-ended question to investigate the rationale for paying attention to the app permissions based on our participants' perspectives.

## 4. Statistical and Descriptive Analysis

In this section, we enhance our results by including a few statistical analyses, as indicated in Figure 1. Such analyses would help to investigate the relationships by relying on the obtained quantitative data. It also helps to determine the significant results within the different demographic data.

### 4.1 Measuring the internal consistency of the study items

Cronbach's alpha is used to measure the internal consistency, and how closely related a set of items are as a group [55]. The Cronbach's alpha is considered as one of the most important and pervasive statistics in research involving test construction and use [48]. The coefficient value should be above 0.70 for a reliable scale. [49]. In our study, we consider the permissions request as one focus area and we wanted to determine the reliability of the measurement value. Our analysis indicated that we exceeded the satisfactory level of construct validity and internal consistency (i.e., $\geq 0.7$) by having 0.905, which is considered as a strong coefficient value [48].

### 4.2 Measuring the correlations of the study items

To further assess the relationship between the permissions request items (i.e., strength and direction of the linear relationship), we conducted Chi-square test to whether there is a statistical significant correlation or not among the obtained data. It also helps to understand the direction of the relationship [45]. Such an investigation can help us to understand when one access permission type changes in value, the other access permission type tends to change in a specific direction. As illustrated in Table 3, we found that all the investigated permissions have a positive and statistical significant correlations at 0.01 level. For example, there is a statistically significant correlation between access to device storage and device camera and the chances of observing the obtained correlation (0.792) through random error are less than 0.01. The correlation results ranged from a strong (e.g., 79%: permission to storage and permission to access device camera) to a moderate (e.g., 49%: respond to pop-up window and permission to access device location) relationship. To further elaborate on one example, the correlation indicates that when the score of permission to access storage increases, we expect (i.e., more likely) the score of accessing device camera to increase positively (it is not a causality relationship). Additionally, our analysis revealed that all the tested items were significantly correlated at 0.01 level.

Table 3. Correlation for the Permissions Requests

| Permission type | Respond to pop-up window | Permission to storage | Permission to Camera | Permission to contacts | Permission to audio recording | Permission to location |
|---|---|---|---|---|---|---|
| Respond to pop-up window | 1 | | | | | |
| Permission to access device storage | .486** | 1 | | | | |
| Permission to access device Camera | .503** | .792** | 1 | | | |
| Permission to access device contacts | .540** | .602** | .770** | 1 | | |
| Permission to access device microphone | .502** | .526** | .694** | .730** | 1 | |
| Permission to access device location | .486** | .581** | .639** | .641** | .717** | 1 |

**. Correlation is significant at the 0.01 level (2-tailed).

## 4.3 Participants reaction about permissions requests

The security awareness for users towards using smartphones, in general, can be measured by understanding how users deal with four technological aspects, namely applications, communication and browsing, channels, and device [12]. Each aspect has a few attack scenarios that can be further investigated [12]. For example, access permissions and users reaction towards fake links were classified as examples of applications security. In our study, we explicitly attempted to exploit the human vulnerability (i.e., lack of security awareness) to exploit specific targets (e.g., device contacts) to gain an understanding of how the users' reacted (i.e., denied or granted access). Based on the type of the app that we developed for our simulation, no access permissions were mandatory for the app to be functioning. Denying these permissions would create no issue for app usage, and granting access could put users data at risks (e.g., allow third-party to access sensitive data). Overall, we requested 630 permissions (i.e., six per participants) in simulation app and we recorded their reactions without reaching the requested resources (including the pop-up window). We found that 296 access permissions (i.e., 47%) were granted to our simulation app, and 334 access permissions (i.e., 53%) were denied. Whilst 33 participants out of 105 (31.4%) had denied all the requested permissions, 77 participants out of 105 (73.3%) had prevented at least one permission, 28 participants out of 105 (26.7%) had responded to grant all the requested permissions for our simulation app. To be specific, we examined our participants' behaviors when they face a phishing link. We presented a pop-up window that offer them with free exercises which were not available in the app. We found that 47% have granted our request and 53% denied it. Additionally, we requested our participants to grant our simulation app unreasonable permissions to their device software, and hardware, namely, access to the device storage, contacts, camera, microphone, and location. Surprisingly, our findings indicated that 47.5% (± 2.5) of our participants have granted our simulation app access to the requested permissions. Figure 5 presents our participants reactions towards the requested permissions along with the mean and Standard Deviation (SD). Our findings also indicated that granting the app to access the device storage and location were the highest among other permissions (i.e., 50% each).

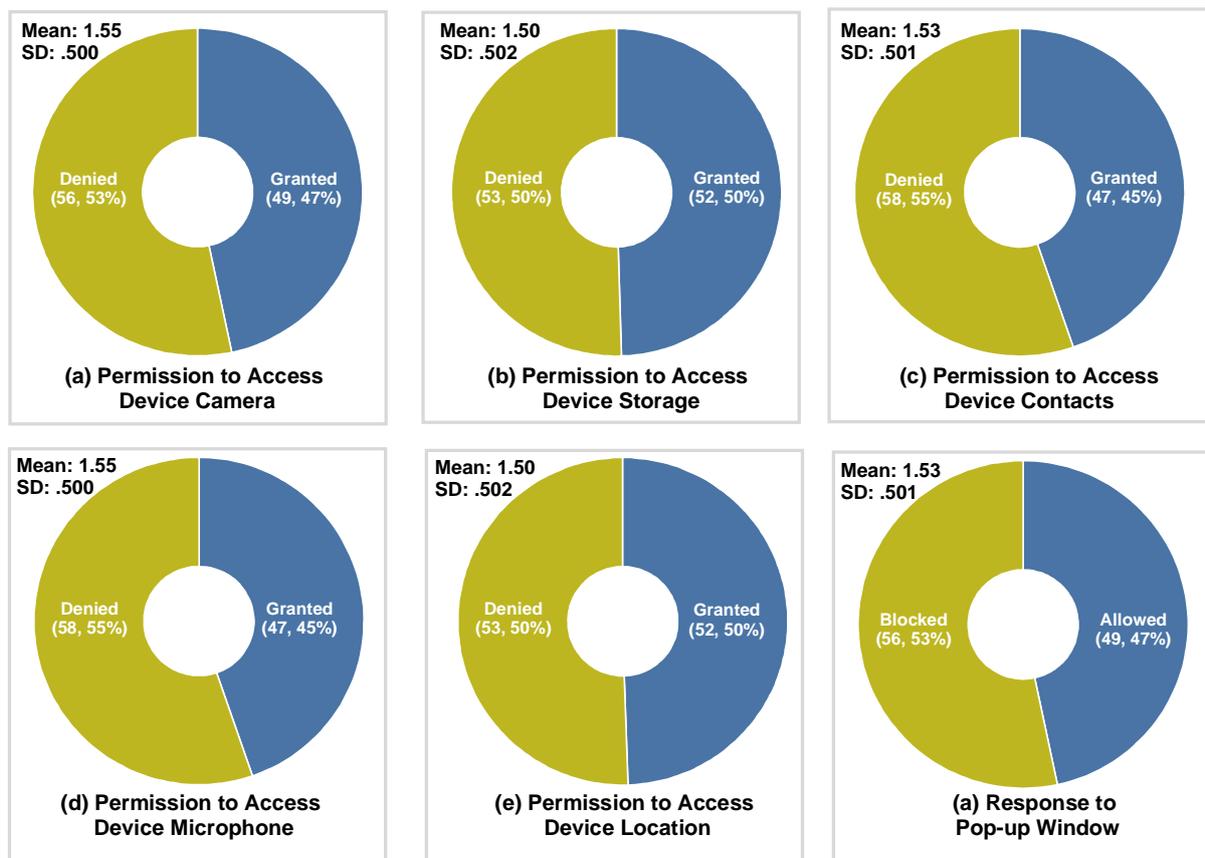

Figure 5. Participants' Reactions for the Requested Access Permissions

## 4.4 Relationship between demographic characteristics and participants reactions to access permissions

As indicated in (Section 3.3 Phase 3 – Perform Data Analysis), we performed Mann–Whitney U test for gender since we were comparing two independent samples (i.e., male and female), and Kruskall-Wallis H test for more than two independent samples (e.g., IT knowledge level, age group, etc.). These statistical tests helped us to determine whether there were statistically significant differences in regards to our participants' reactions to the requested access permissions among the defined groups of users, as in Table 4. The findings provides fine-grained analysis of the security reactions to access permissions for each specific group. For example, which access permissions received the highest allowance by female participants, which group of users are more likely to click on a phishing link, etc. We further investigated the significant differences for the groups whenever applicable (i.e., whenever p-value < .05). We used the Mann Whitney U test to compare the median differences between the overall extents [9, 22]. For each demographic data, we tested the null hypothesis (i.e., *H0: there is no significant difference*) against the alternative hypothesis (i.e., *H1: there is a significant difference*), whereas *μ1, μ2, …, μk* refers to population means.

Table 4. Variables and Corresponding Codes in SPSS for the Attack-Simulation Study

| Simulation app Permissions | Code | Gender | Code | IT Knowledge Level | Code | Age Group | Code | Formal Education | Code |
|---|---|---|---|---|---|---|---|---|---|
| *Granted (Allowed)* | 1 | *Male* | 1 | *Little or no knowledge* | 1 | *18 – 29 young adult* | 1 | *High school or less* | 1 |
| | | | | | | | | *Diploma* | 2 |
| *Denied (Blocked)* | 2 | *Female* | 2 | *Moderate knowledge* | 2 | *30 – 49 adult* | 2 | *Bachelor degree* | 3 |
| | | | | *Advanced knowledge* | 3 | *Above 50 senior* | 3 | *Master's or PhD* | 4 |

### A. Access Permissions Based on Gender

To understand and compare how male (n=67) and female (n=38) participants reacted with the requested access permissions, we performed a statistical test (i.e., **Mann–Whitney U test**) to show if there is any significant difference. The overall result indicated that there was no significant difference between male and female (u = 1073.00, z= -1.371, p-value= .170) as in Table 5. However, we ran the same test to determine differences in each access permission between male and female. As in Table 6, we found that access permission to device storage is statistically significant (u = 915.00, z= -2.757, p-value= .006). The mean rank for female = 62.42 which is higher than the mean rank for male = 47.66 indicating that female have a higher attention to granting our simulation app access to the device storage. Thus, we concluded that both male and female in our sample have reacted equally to the requested access permissions excluding accessing device storage.

Table 5. Differences towards Permissions Requests Based on the Characteristics of Study Respondents

| Demography Data Category | Identified groups | N (%) | p-value |
|---|---|---|---|
| Gender | Male | 67 (64%) | .170 |
| | Female | 38 (36%) | |
| IT Knowledge Level | Little or no knowledge | 48 (46%) | <.01 |
| | Moderate knowledge | 42 (40%) | |
| | Advanced knowledge | 15 (14%) | |
| Age Group | 18 – 29 young adult | 47 (45%) | <.01 |
| | 30 – 49 adult | 53 (50%) | |
| | Above 50 senior | 5 (5%) | |
| Formal Education | High school or less | 11 (10%) | .026 |
| | Diploma | 14 (13%) | |
| | Bachelor degree | 53 (50%) | |
| | Master's or PhD | 27 (26%) | |

Table 6. Results for Mann-Whitney U test for Gender Sample

| Requested Permission | Respond To Pop-up Window | Storage Permission | Camera Permission | Contacts Permission | Location Permission | Microphone Permission |
|---|---|---|---|---|---|---|
| Mann-Whitney U test | 1182.00 | 915.00 | 1077.00 | 1220.00 | 1177.50 | 1220.50 |
| Z | -.702 | -2.757 | -1.512 | -.410 | -.735 | -.410 |
| P value | .483 | .006 | .130 | .682 | .462 | .682 |

### B. Access Permissions Based on IT Knowledge Level

We conducted the Kruskall-Wallis H test to investigate the significance of granting or denying access permissions based on the participants' IT knowledge (i.e., Little or no knowledge, Moderate knowledge, and Advanced knowledge), as in Table 5. Our findings revealed that reacting to access permissions is differed significantly (p<0.01). For the Moderate IT knowledge mean rank = 63.21 which is less than Advanced IT knowledge mean rank = 61.90 and less than Little or no knowledge mean rank = 41.28. Specifically, significant differences were found (using post-hoc Mann-Whitney U Test) between Little or no knowledge group compared with Moderate knowledge group (p =.001). Whilst we found that there is no statistically significant difference on the security awareness between Moderate knowledge group, and Advanced knowledge group (p=1.000), and there is no statistically significant difference on the security awareness between Little or no knowledge group, and Advanced knowledge group (p=.056). Therefore, we reject H0 and concluded that IT knowledge level had an impact on their decisions to grant or deny access permissions for our participants.

### C. Access Permissions Based on Age Group

We conducted Kruskall-Wallis H test to determine any significant difference among the three age groups (i.e., n=47; young adult, n=53; adult, n=5; senior) regarding to how they reacted to the requested access permissions. Our simulation app was developed to help the participants to do exercises as we were advertising for our study. We believe that we were not attracting as many seniors as we should. This could be the main reason that we had a very low number of Senior group (5 out of 105). As in Table 5, our findings suggested that there is a statistically significant (p=0.001) when reacting to the requested access permission for the three age groups. Adult group (mean rank = 62.98) were less than Young Adult group (mean rank = 45.49), were less than Senior group (mean rank = 21.30). Whilst we found that there is no statistically significant difference between Senior group and Young Adult group (p= .247), our results revealed that there is a statistically significant difference between Senior group and Adult group (p= .008). We also found that there is a statistically significant difference between Young Adult group and Adult group (p= .010). Therefore, we reject H0: $\mu_1 = \mu_2 = \mu_3$ and conclude that age did have an impact on the participants when reacting with the requested access permissions.

### D. Access Permissions Based on Level of Formal Education

We also wanted to investigate respondents' differences by considering the impact of the level of formal education on our participants. We conducted Kruskall-Wallis H test on the four groups, which we identified in Table 5. Our results indicated that there is an evidence (p=0.026) that reacting to requested permissions of those with a postgraduate qualification (mean rank= 60.87) was lower, than those with an education level of Bachelor (mean rank= 56.15), less than those who are having Diploma (mean rank= 40.14), and less than those with High School or less (mean rank= 34.86). We went further to understand the difference within the four level of education through using post-hoc Mann-Whitney U test. We found that High school or less group and Bachelor group have a statistically significant difference in granting or denying access permissions (p= .030). We also observed that High school or less group and Postgraduate group have statistically significant difference (p= .014). Further, we noticed a statistically significant difference between Postgraduate group and Diploma group (p= .034). Our conclusion based on the obtained results suggested to reject H0: $\mu_1 = \mu_2 = \mu_3 = \mu_4$. Thus, the level of formal education did have an impact on the participants when dealing with access permissions.

## 4.5 Time spent of reviewing privacy policy by our participants

App privacy policy is a legal statement that regulates the engagement with users. It presents to the users how and when their personal data would be collected, retained, or shared with a third party. It also explains what resources the app is requesting, and for what purpose including access permissions, and the financial obligations for the app usage [50]. Some installed app can do an activity that is risky for user without they being realized [27]. Hence, the privacy policy should be carefully read by users before agreeing. In our simulation app and as most mobile apps do, we presented the privacy policy before the installation for our participants, as in Figure 3 (a). Our aim was to investigate how the participants would react and how much time they spent on reading it. To calculate the time, we recorded the time once the privacy policy was presented and the time the participants clicked on the "Continue" button. The average time spent on reviewing the privacy policy for all participants was 8.02 ms to review about 500 words, which requires at least a minute.

Due to the fact that IT knowledge can make a difference in regards to dealing with mobile apps, we further investigated reviewing the privacy policy behavior by our participants by looking into their IT knowledge level. As in Table 7, we found that participants with advanced IT knowledge (n=15) spent an average time 19.61 ms. While participants with moderate IT knowledge (n=42) spent an average time 7.26 ms, and the participants with little or no knowledge on IT (n=48) spent an average time 5.06 ms. The participant identified as **P96** who spent 5 seconds in reviewing privacy policy wrote "*I don't pay attention to the privacy policy except when the duties and financial responsibility towards the app are mentioned. For example, the policy of cancelling the subscription if the app requires payment or financial fees*".

Table 7. Total and Average Time Spent on Reviewing Privacy Policy

| Participants Group | Total time spent | Average |
|---|---|---|
| All study participants (n=105) | 842 seconds | 8.02 ms |
| Participants with little or no IT knowledge (n=48) | 243 seconds | 5.06 ms |
| Participants with moderate IT knowledge (n= 42) | 305 seconds | 7.26 ms |
| Participants with advanced IT knowledge (n= 15) | 294 seconds | 19.60 ms |

### 4.6 Preferred authentication method for our participants

User authentication in our context refers to how the participant wants to log into the simulation app and what method can be used to prove that s/he the legitimate app user. In our previous study [3], we surveyed 101 mHealth apps users to investigate the desired security features they want to be employed. We found that users were having different opinions about the authentication method (e.g., employing biometric authentication, interactive authentication, direct access, etc.). Hence, in our simulation app, we provided a variety of options for authentication (ranging from weak, moderate or robust security) to investigate users selection. We asked our participants to customize the simulation app by selecting their preferred authentication method. We presented five common authentication methods, namely, none (i.e., without authentication), pattern, Personal Identification Number (PIN), username and password, and Two-factor authentication (2FA). As in Figure 6, 38 (36%) of our participants preferred no authentication method. Using Personal Identification Number (PIN) to access the app was selected by 20 participants (i.e., 19%). Using pattern was the preferred option for 13 participants (12%). 23 participants (22%) preferred to have a username and password to log in to the app, and only 11 participants (10%) preferred 2FA.

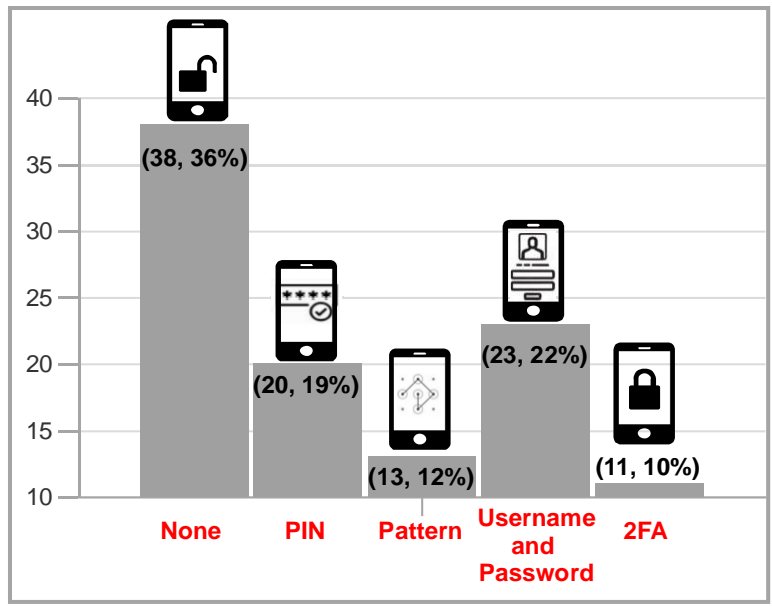

Figure 6. Participants Favorable Authentication Method

### 4.7 Participants interest in reporting security issue and receiving security advice

In our simulation app, we wanted to investigate the participants' behaviors towards reporting security issues within the simulation app in case they noticed. As in Figure 7, our findings indicate that only 36 (34%) participants wanted to report security issue and 69 (66%) participants did not want to report any security issue. However, we are unsure whether these participants actually noticed an issue or not. One possible reason is that participants ignored what they have seen and did not want to bother about these issues. In fact, **P97** wrote "*When I see a security issue with the app, I just uninstall it*". For further analysis, we looked into the level of IT knowledge for the participants who wanted to report security issue(s) within the simulation app. We found that 9 participants with advanced IT knowledge (n=15), 12 participants with moderate IT knowledge (n=42), and 15 participants with little or no IT knowledge (n=48).

We also asked our participants whether they want to receive frequent security advices or not. Our results indicated that 76 (72%) participants accepted getting security advices and 29 (28%) participants refused this idea, as in Figure 7. For further analysis, we looked into the level of IT knowledge for the participants who decided on not to receive any security advice. Interestingly, the participants were from different levels of IT knowledge. 4 participants with advanced IT knowledge (n=15), 13 participants with moderate IT knowledge (n=42), and 12 participants with little or no IT knowledge (n=48).

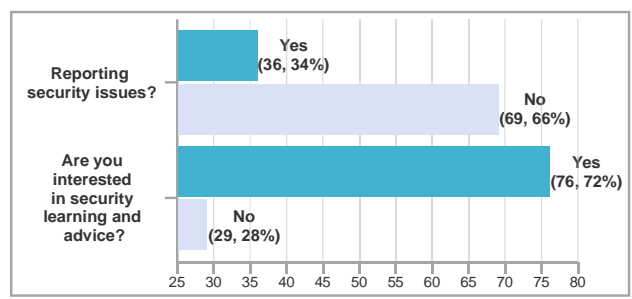

Figure 7. Participants Reactions for Reporting Security Issues and their Interest in Security Learning

## 5. Reasons for paying attention to the app permissions

In this section, we provide our analysis for the open-ended question that we included in the exit survey, as detailed in Section 2.4.1. We aimed to identify the reasons that make our participants pay attention to the requested app permissions based on their views. Based on the obtained responses, we used the qualitative descriptive analysis (Section 3.3 – Phase 3 – Perform Data Analysis) to create the conceptualization, as in Figure 8. Whilst a few participants perceived the requested permissions positively (Section 4.2.1), the majority had a negative views about the requested permissions as in (Section 7.3.2.2 – Section 7.3.2.5). Violating/losing privacy were explicitly mentioned by some participants, e.g., **P46**, **P47**, **P51**, **P54**, **P68**, **P82**, **P83**, **P88**, **P89**, **P92**, and **P98**. Four participants **P42**, **P47**, **P66**, and **P71** indicated that the requestion permissions were unnecessary. **P71** stated, *"[the app] should not access my contacts, photos, or location because the app does not need access to my information"*. We provided cherry-picked examples for each theme to support our findings in Figure 7. Each theme is discussed in the following sections.

### 5.1 App permissions provide a better usage

The ideal permission for our simulation apps which would make sense to request is accessing fitness data (Health app in iOS). This permission would help to retrieve the data (e.g., steps count, burnt calories, etc.) into our app and perform the required analysis for users. However, we did not request this permission purposefully. Instead, we requested permissions to camera, contacts, microphone, location, photos that would make no logic to be granted. Five participants **P30**, **P45**, **P65**, **P82**, **P87** considered the requested permissions were necessary for the app functionality. The participants assumed that the simulation app was requesting permissions to perform relevant tasks. **P30** mentioned, *"I assume the app needs permissions to help me to have better usage"*, and **P87** wrote, *"It is ok to grant the app to access my location because it is already the case with all other apps"*.

### 5.2 Fear of data leakage

Data leakage can be defined as the unauthorized distribution for any confidential or sensitive users data [10]. Whilst app permissions is not the only method for data leakage, providing personal information when registering for the app could potentially leaked without app's provider/developer awareness [23]. We are reporting the participants' feedback regarding the reasons for granting or denying app permissions. Our findings revealed that 16 participants **P34**, **P36**, **P46**, **P51**, **P53**, **P54**, **P65**, **P68**, **P71** - **P74**, **P84**, **P89**, **P92**, **P98** were having fear of data leakage for either their personal information (e.g., contact number, email address) or device data (e.g., contacts, photos, location). Four participants **P34**, **P51**, **P53**, and **P89** specified that app permissions would permit the app's provider/developer to access their private data. **P89** wrote, *"App permission, based on my knowledge, allow the developer to access my data in the device"*, **P51** stated, *"Giving access to your photos and contacts when it is not necessary can lead to giving the app access of transferring sensitive content to the developers"*, and **P53** indicated *"when I have sensitive data on my phone, [access permissions] my take sms, contacts, or mails to do some things I don't like."* Four participants, **P51**, **P68**, **P72**, **P92** indicated that they want to protect and maintain their private data. **P51** added, *"...I value my privacy and everything that is on my phone"*, and **P68** wrote, *"To maintain the privacy of my data"*, and **P92** wrote, *"Protecting my personal information and data"*. Three participants **P54**, **P65**, and **P71** mentioned that access permissions would lead to intentionally releasing or making the information publicly available to others. **P71** stated, *"The impact can be sharing data publicly such as the app that store the contacts and make it available to others"*, and **P65** wrote, *"Protect my data from being spread out on the internet"*. Five participants indicated giving access permissions can make their private data vulnerable to data leakage. **P84** wrote, *"I pay attention because I have private pictures on my phone and I don't want it to be stolen"*, **P73** wrote, *"It may be leaked my location information and mobile number"*, and **P74** wrote, *"may lose my personal information"*.

### 5.3 Fear of their devices being hacked

Access permissions can be the role of attack surface that could be exploited. A compromised app would allow attackers to exploit whatever the app is already connected with including camera, microphone [20]. Six participants **P42**, **P46**, **P49**, **P86**, **P95**, **P102** expressed their fears of giving more permissions to our simulation app may lead to make their devices insecure. The participants indicated that doing so may make their devices vulnerable for malicious activities such as spying (**P42**, **P46**), or their devices may become hackable (**P49**, **P86**, **P95**, **P102**). Specifically **R42** stated *"It will be a back door that will open vulnerabilities to be used to harm the device and its data"*, and **R46** wrote, *"Spy on my data and activity"*. Furthermore, **P86** indicated *"Sometimes permission can be hackable if it not well secured"*, and **P49** mentioned *"To protect myself from hacks"*.

### 5.4 Excessive permission triggers a lack of trust

As indicated in Section 2.4.1, we developed our simulation app and we hosted it on our storage. Since participants had to download from unknown source, most likely this process gave an impression that our app was suspicious (no reviews or rating available to view). When we asked our participants the reasons of granting or denying the requested permissions, Four participants **P66**, **P73**, **P90,** and **P105** indicated that the simulation app is not trusted to be granted the requested permissions. For example, **P73** stated, *"I will be unhappy about the more permission than needs and don't trust this app using, so that I will stop the app from accessing my information"*. **R66** found that app permissions should not be requested. **R66** wrote, *"I don't let it go easily. I denied the requested permissions because the app does not need them. Only malicious app would ask for these access privileges."*

### 5.5 Fear of using data for objectionable processing purposes

Gathering users' data (e.g., contacts, location, etc.) through app permissions can be done easily. However, it becomes a problematic in case the app providers share that data without declaration or consent from users. This data processing would be objected by users if they were asked in advance. Four participants pointed out this concern and indicated that app permissions would gather data to make profits (**P46, P47, P49, R68**). For instance, **P68** stated *"Saving the data and distribute it to private parties. Some companies enable some apps to sell their clients' data"*, **P47** wrote *"Transforming users into products, by selling their information"*, and **P49** indicated *"It collects a huge data from the user, and the developer usually sell that data to someone else"*. In addition, two participants reported that the purpose of requesting more permissions is sending advertisements (**P30, P73**). **P73** wrote, *"It may continue to sent my mobile some useless messages latterly"*, and **P30** stated *"Sending ads to me"*.

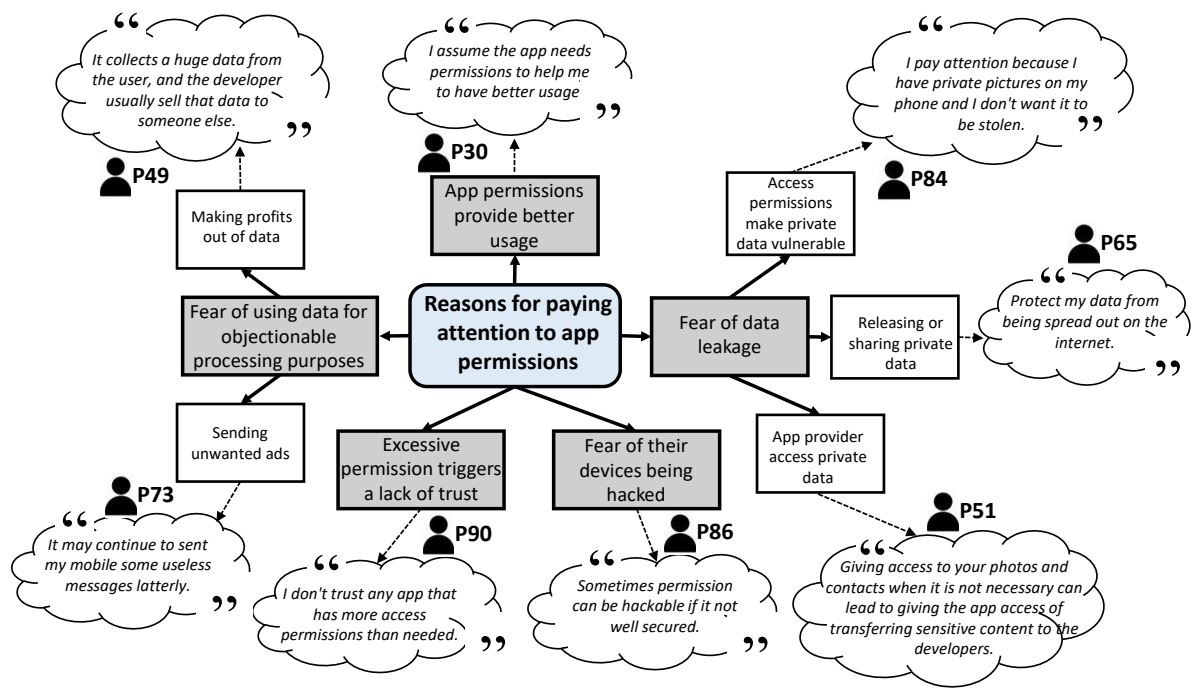

Figure 8. Participants' Views on the Reasons for Paying Attention to Access Permissions with Some Cherry-Picked Example for Each Theme.

## 6. Discussion

We now discuss the results that highlight the core findings of this study, based on methodological details in Figure 1, and outline the potential future work. Table 8 guides the discussion and presents a summary of the key findings. The discussion highlights reviewing privacy policy (Section 6.1) followed by discussing the authentication method for the simulation app (Section 6.2). We then discuss the results of participants' reactions about requesting access permissions (Section 6.3), and finally the reasons for paying attention to the requested permissions is discussed in Section 6.4.

Table 8. Taxonomical Classification of the Core Findings for Attack Simulation Approach Study

| Measuring the Security Awareness of Users about Using mHealth Apps: Attack Simulation Approach | | | | |
|---|---|---|---|---|
| Gender Classification | Age Group | Formal Education | IT Knowledge Level | Country |
| • 64% Male<br>• 36% Female | • 45% 18 – 29<br>• 50% 30 – 49<br>• 5% above 50 | • 10% High school or less<br>• 13% Diploma<br>• 50% Bachelor's<br>• 26% Postgraduate level | • 46% little/no knowledge<br>• 40% moderate knowledge<br>• 14% advanced knowledge | • 39% Saudi Arabia<br>• 19% Australia<br>• 18% India<br>• 10% China<br>• 14% Others |
| Users security reactions when facing potential security threats | | | | |
| Reviewing Privacy Policy | Selecting authentication method for the app | Reactions about requesting access permissions and phishing | | Reporting security issue | Interest in receiving security learning and advice |
| • All participants (n=105) spent 14 minutes and 03 seconds.<br>• Average of 8.02 ms per participant | • 36% None<br>• 19% PIN<br>• 12% Pattern<br>• 22% Username/ password<br>• 10% 2FA | Permission Type | Granted | Denied | • 34% Yes<br>• 66% No | • 72% Yes<br>• 28% No |
| | | Pop-up window | 47% | 53% | | |
| | | Device storage | 50% | 50% | | |
| | | Contacts | 47% | 53% | | |
| | | Camera | 45% | 55% | | |
| | | Microphone | 45% | 55% | | |
| | | Location | 50% | 50% | | |
| Reasons for paying attention to the requested access permissions | | | | |
| • App permissions provide a better usage<br>• Fear of data leakage<br>    - Access permissions make private data vulnerable<br>    - Releasing or sharing private data<br>    - App provider access private data<br>• Fear of their devices being hacked<br>• Excessive permission triggers a lack of trust<br>• Fear of using data for objectionable processing purposes<br>    - Making profit out of data<br>    - Sending unwanted advertisements | | | | |

## 6.1 Time Spent on Privacy Policy

Reviewing privacy policy is a vital document that allow apps developers to inform users about their data collection practices [37]. Some mobile apps do not provide clear transparency about exactly what data is being shared, with whom, and for what purposes [40]. App developers may share users' data with third parties for advertising purposes, or data can be leaked without developers having any thought about how that occurred [42]. The results showed that users spent on average 8.02 ms to review the presented privacy policy (about 500 words), which is not adequate to carefully review it, as in Table 7. Only nine participants (8%) spent more than 20 seconds reviewing the privacy policy, presented in Figure 3 (a). Interestingly, three participants spent rational time reviewing it (i.e., a minute and 55 seconds, a minute and 13 seconds, a minute and 3 seconds, respectively). Most of the users skip reading through the privacy policy and look for required action (ticking the box) to install the app. The finding of this is consistent with previous research such as [37, 40, 42]. It should be noted that the entire time which our participants spent on the privacy policy page does not mean that participants were reviewing it. The time was calculated from the time that privacy policy page loaded until the participants clicked on continue button. Therefore, future research could consider employing accurate mechanisms such as eye-tracking (requires accessing the device camera) to record the time which participants are actually spending. Whilst there are several studies that indicated that there is a lack of clarity and transparency in presenting such policies for mHealth apps [36, 37, 40], and recommendations to improve the privacy policy for users [1, 24, 29, 33], further work can be done to examine users reactions when presenting the privacy policy through using innovative methods including summarizing the main points, using visualization, or a displaying it in video. This approach would be effective to encourage users, especially those who have little or no IT knowledge, to spend more time to understand what data, and how their data are being handled. Exit survey or semi-structured interviews can be considered to allow participants to share their thoughts. Such a study would help to understand how users review privacy policy and their understanding of the associated risks of using the apps.

## 6.2 Authentication Method for the Simulation App

Each type of mobile apps necessities access to the device hardware (e.g., camera, audio recording, etc.) or software (e.g., contacts, images, etc.) to be functioning. Certain apps which deal with sensitive data, require security measures and security awareness from users as well. Authentication mechanisms is an important measure that is provided for users to provide a layer of protection at the application level. Lack of users authentication is considered as serious risk for users' data [24]. We aimed to assess participants' favorable authentication method for the simulation app. We offered them different authentication methods to select from based on their assessment. 67 out of 105 participants (i.e., 63%) believed that our simulation app needs an authentication method, such as PIN, pattern to log in, as illustrated in Figure 2, and Table 8. In contrast, 38 out of 105 participants (36%) selected to access the app without user credentials. We further examined the "none" choice among the different groups of users based on their IT knowledge. We found 18 out of 48 (37%) participants with little or no knowledge in IT selected no authentication. 15 out of 42 (36%) participants with moderate IT knowledge and five out of 15 (33%) participants with advanced IT knowledge selected no authentication. The results indicated that about one-third of each IT level groups preferred no authentication for the simulation app. Unlike our findings in section 4.4 (b), which indicated that IT knowledge level has a statistical significant difference on app permissions, we concluded that IT knowledge background has no impact on participants' selections for app authentication method. It is more likely that our participants found that the app is for fitness (based on the study invitation); and hence, the app does not seem to be having sensitive data. Further work needs to consider adopting an app that contains more personal data. This would help to investigate the differences of users' security reactions in regards to the authentication method. It would be even more insightful to include qualitative data collection (e.g., interviews) to understand the reasons for the participants' selections.

## 6.3 Access Permissions for the Simulation App

App permissions are designed to ensure that mobile apps access the required resources to work properly [30]. However, requesting access permission beyond the app's purpose of the app is considered malicious access permission. In our simulation app, as in Table 8, we requested five permissions to access the participants' hardware/software of their devices to examine their reactions. The requested permissions appeared randomly for the participants at the time of installation and during the runtime of the app. We found that 52 participants (50%) granted the simulation app access to their location. Table 9 presents further details about the participants who granted or denied accessing to their location based on the IT knowledge. **R87** wrote *"It is ok to grant the app to access my location because it is already the case with all other apps"*, (a female participant with a bachelor degree, and little or no knowledge in IT). Most likely that users have trust in giving some permissions, such as location or specific hardware (e.g., microphone). This could lead to further work to investigate what other access permissions that users trust, and the reasons that make them trust.

The findings in Section 4.4 (a) indicated that there was no statistical significant difference for access permissions request based on the gender, as in Table 6. However, we further found that 53 participants (50%) denied the simulation app access to the storage of their device. Table 9 presents further details about the participants who granted or denied accessing to their device storage based on the IT knowledge. To provide further analysis, we looked into the denied permissions based on gender, as in Table 10. We found that females have a higher concern about granting access to their devices. Due to the fact that device storage store their private photos which cannot be shared, accessing device storage received the highest denial by 26 females (68%). This point confirms our results in Table 5 which indicates a statistical significant difference between female and male participants in regards to accessing device storage. Accessing to the device location had the lowest denial by 21 females (55%). **R71** wrote, *"[The app] should not access my contacts, photos, or location because the app does not need access to my information"*, (a female participant with a bachelor degree, and moderate IT knowledge). On the other hand, we found that out of 67 males, 27 (40%) denied access to device storage which is the lowest denial. The highest denial from males were contact and microphone (i.e., 36 males (54%) for each permission request). Table 10 presents the numbers of male and female who granted/denied the access permissions. **R51** wrote, *"giving access to your photos and contacts when it is not necessary can lead to giving the app access of transferring sensitive content to the developers"*, (a male participant with a bachelor degree, and moderate IT knowledge). In our conclusion, females have a higher concern about requesting access to the device storage. Device storage store their private photos that cannot be shared others.

Table 9. Participants Responses to Requesting Access Permission to their Devices' Locations and Storage based on IT Knowledge Level

| Access Permission Type | Options | Little or no IT knowledge (n=48) | Moderate IT knowledge (n=42) | Advanced IT knowledge (n=15) |
|---|---|---|---|---|
| Access to Location | Granted, 52 (50%) | 31 (64%) | 16 (38%) | 5 (33%) |
| | Denied, 53 (50%) | 17 (36%) | 26 (62%) | 10 (67%) |
| Access to Device Storage | Granted, 52 (50%) | 29 (60%) | 16 (38%) | 7 (47%) |
| | Denied, 53 (50%) | 19 (40%) | 26 (62%) | 8 (53%) |

Table 10. Participants Responses to deny the Requested Access Permission based on Gender

| Denied access permissions requests based on gender | Device storage | Camera | Contacts | Microphone | Location | pop-up Window |
|---|---|---|---|---|---|---|
| Female (n=38) | 26 (68%) | 24 (63%) | 22 (58%) | 22 (58%) | 21 (55%) | 22 (58%) |
| Male (n=67) | 27 (40%) | 32 (48%) | 36 (54%) | 36 (54%) | 32 (48%) | 34 (51%) |

### 6.4 Paying attention to the requested permissions

One of the most common approach for developers to make profit out of their apps is embedding an advertising library, which displays ads to the users when using the app. Nevertheless, these third party advertisement libraries are granted the same permissions as the developers, and may share or on-sell the information they collect with other entities in the mobile ecosystem [37]. At the same time, mobile apps can be vulnerable to security breaches, especially when the app developed through poor security practices [7]. The simulation app which the study used requested access to device resources (six types of resources as presented in Section 4.1.3, and Table 8). The simulation app purposefully lacks transparency (i.e., no clarifications why these permissions were requested) to examine and record the participants' reactions. All requested permissions popped up to the participants soon after the app installation or when they were using the app. Almost 50% of our participants granted the simulation app permissions. Interestingly, a few participants claimed that they paid attention to app permissions because they were worried about their privacy. However, we found contradiction with their responses in the exit survey. For example, **R34**, and **R73** (both are males with moderate IT knowledge) fully granted our simulation app the requested permissions. **R34** wrote, "*To not get permissions to private images*", and **R73** wrote, "*I will be unhappy about the more permission than needs and don't trust this app using, so that I will stop the app from accessing my information*". The results also indicated that users can be easily fall into the ambush of sharing their data and devices. This triggers the alarm that users need to pay careful attention when installing and using mobile apps. Further work can be done using similar approach with allowing the participants to share their views about granting or denying permissions after every single request. In addition, a summary report can be shared with each participants as soon as s/he completes the study. Such a summary should explain what mistakes the participant has done with their potential impacts, and what could have been done better when facing these access permissions. Consequently, this would help to increase the security awareness of the users before thinking to grant any app a permission.

### 7. Threats to Validity

We now discuss some threats to the validity that represent potential limitations that might impact the finding of the study. Validity threats are a results of certain assumptions or constraints that need to be highlighted and ideally eliminated as part of future research.

### 7.1 Internal validity

Internal validity refers to the factors that could have impacted data collection and analysis processes negatively. The study collected data through a simulation app that we developed. It presented some real-life security scenarios that any users face during mobile apps usage. However, the study findings might get affected since we had to explicitly indicate to the participants that we were investigating their security awareness. Initially our plan was to hide the aim of the study by mentioning that we are going to investigate participants' habits and behaviors towards mHealth apps. Yet, due to the limited-disclosure of the study aim, we were not able to get the ethical approval. Indeed, the application of our study was elevated to a high-risk review, and it took so many reviews until we addressed the concerns of the ethics committee (i.e., explaining the aim of the study for participants in advance). Another threat can be analyzing qualitative data of the exit survey, which may lead to misinterpretation of participants' responses. To overcome this threat, the initial coding of the data was done by one person (i.e., the first author) followed by evaluation and finalization of the codes by another researcher).

### 7.2 Construct validity

Construct validity refers to the extent to which the used instrument measures what it is supposed to be measured. Due to the limited time to finish the study, the app was available to download from our storage. Thus, we found collecting data is a bit challenging since we were trying to convince participants to download an app using .APK file without referring to the apps market. Participants were skeptical because most of them received a warning message from their devices to not download the app as it can be a malicious app. Our study could be also impacted by the limited security scenarios that we already measured. Some security scenarios (e.g., password strength, password changing habits) were questionable by the ethics committee; hence, we had to drop them to get the study done. The other limitation is that our simulation app is only compatible with Android devices. The given time to finish the research has limited us to configure the app with iOS devices.

### 7.3 External validity

External validity refers to applying generalization of the study results. The findings of our study cannot be generalized since we have limited number of some categories (e.g., a few participants above the age of 50). Thus, we plan to further extend this study by including more security scenarios, other research method (e.g., semi-structured interviews, think aloud, etc.), and more open-ended questions. In addition, we plan to reach a large number of audience geographically distributed (i.e., include participants with spoken languages other than English). This would help us to increase the number of participants and diversity of data collection.

### 7.4 Conclusion validity

Conclusion validity refers to the factors that affect the credibility of the study conclusions. To address this threat, the first three authors were continuously involved in the interview instrument development and data analysis process. However, all other authors occasionally (consent meetings) reviewed the data and provided their feedback to tackle the conflicts that appeared during the data collection and analysis process. Finally, brainstorming sessions were conducted, and all the authors participated in discussing the study findings and drawing conclusions.

## 8. Conclusions and Future Work

Mobile computing as the backbone of context-sensitive digital servicing is revolutionizing the smart and mobile health across various domains of medical practices that range from clinical apps, to fitness trackers to diet and nutrition decision support [36]. Despite the offered benefits such as autonomy, efficiency, and pervasiveness of healthcare services, mHealth apps face some challenges pertaining to security and privacy of health critical data. In addition to developing secure mHealth apps and optimizing the methods and techniques for secure storage, processing, and transmission of data, there is a need to understand and enhance the security awareness of the users. Security understanding of mHealth app users relates to their knowledge and behavior (often translating into actions) to ensure secure usage. In addition to the technical measures that ensure security of mHealth apps, enhancing the security awareness of mobile users is critical due to the high number of security threats and the numerous malicious activities. Such an awareness would help to secure confidential data stored within the devices. We conducted an empirical study to monitor and analyze the actions and behaviors of mHealth app users – relying on an attack simulation approach – to overcome the inherent bias and limitations of the self-reported surveys. The key findings of our study are:

- mHealth users most often overlook the privacy policies while installing or updating the apps, resulting in issues of apps accessing classified data with user granted permissions. We noted that the average time spent on reviewing privacy policies was 8.02 ms, that is 67% less than bare minimum time to go through the policies (at least a minute to review).
- Users (i.e., 36% of our participants) selected no security method for the simulation app as their favourable method to access the app and 47.5% (± 2.5) of our participants have granted our simulation app access to the unnecessary permissions (e.g., accessing contacts, camera, location, etc.).

- The majority of participants had negative views about the requested access permissions (fear of data leakage, fear of their devices being hacked, excessive permission triggers a lack of trust, fear of using data for objectionable processing purposes).

*Dimensions for potential future work:* This study examined users' security awareness through an attack simulation approach, and it has uncovered some possible directions for future work. It could be further extended to include a larger number of participants with different demographic information. Future studies could also include other security threats which have not been covered in this study (such as strength of passwords, changing passwords, data security at rest and data in transit, etc.). It could be further extended by using the same approach but by also providing a summary report to be shared with each participants as soon as s/he completes the study. Such a summary could explain what mistakes the participant has made and potential impacts of those mistakes, and what could have been done better when facing these access permissions.

*Implications of research:* The study can have the following implications for research and development of secure and usable mobile health solutions.

- Researchers who are interesting in exploring security awareness of users, formulating new hypothesis to be tested for secure mHealth apps, and conducting action-based research on secure mobile computing.
- Developers and stakeholders to understand security specific knowledge and behaviors of user, monitored by our proof of the concept attack simulator and documented guidelines, to develop and provide emerging and futuristic class of mHealth apps that are secure and usable.

## Acknowledgments

We thank Abdulrahman Gharwi for sharing his experience that helped to develop the simulation app.

## References

[1] R. ADHIKARI, D. RICHARDS, AND K. SCOTT. 2014. Security and privacy issues related to the use of mobile health apps. In *Proceedings of the 25th Australasian Conference on Information Systems, ACIS 2014*.

[2] B. ALJEDAANI. 2021. Measuring the Security Awareness of End-Users towards Using Mobile Health Apps: An Attack Simulation Approach [Supplementary Data]. [Online:] https://sites.google.com/view/attack-simulation-approach/home.

[3] BAKHEET ALJEDAANI, AAKASH AHMAD, MANSOOREH ZAHEDI, AND M ALI BABAR. 2021. End-Users' Knowledge and Perception about Security of Mobile Health Apps: A Case Study with Two Saudi Arabian mHealth Providers. *arXiv preprint arXiv:2101.10412*.

[4] BAKHEET ALJEDAANI, AAKASH AHMAD, MANSOOREH ZAHEDI, AND M. ALI BABAR. 2020. Security Awareness of End-Users of Mobile Health Applications: An Empirical Study. In *Proceedings of the MobiQuitous 2020 - 17th EAI International Conference on Mobile and Ubiquitous Systems: Computing, Networking and Services* (Darmstadt, Germany2020), Association for Computing Machinery, 125–136. DOI= http://dx.doi.org/10.1145/3448891.3448952.

[5] FAISAL ALOTAIBI, STEVEN FURNELL, INGO STENGEL, AND MARIA PAPADAKI. 2016. A survey of cyber-security awareness in Saudi Arabia. In *2016 11th International Conference for Internet Technology and Secured Transactions (ICITST)* IEEE, 154-158.

[6] AUDIE A ATIENZA, CHRISTINA ZARCADOOLAS, WENDY VAUGHON, PENELOPE HUGHES, VAISHALI PATEL, WEN-YING SYLVIA CHOU, AND JOY PRITTS. 2015. Consumer attitudes and perceptions on mHealth privacy and security: findings from a mixed-methods study. *Journal of health communication 20*, 6, 673-679.

[7] B. ALJEDAANI, A. AHMAD, M. ZAHEDI, AND M. A. BABAR. 2020. An Empirical Study on Developing Secure Mobile Health Apps: The Developers' Perspective. *27th Asia-Pacific Software Engineering Conference (APSEC), Singapore, Singapore*, 208-217. DOI= http://dx.doi.org/10.1109/APSEC51365.2020.00029.

[8] SUSANNE BARTH, MENNO DT DE JONG, MARIANNE JUNGER, PIETER H HARTEL, AND JANINA C ROPPELT. 2019. Putting the privacy paradox to the test: Online privacy and security behaviors among users with technical knowledge, privacy awareness, and financial resources. *Telematics and Informatics 41*, 55-69.

[9] RUBEN GEERT VAN DEN BERG. November 2020. SPSS Mann-Whitney Test – Simple Example available at https://www.spss-tutorials.com/spss-mann-whitney-test-simple-example/.

[10] YOANN BERTRAND, KARIMA BOUDAOUD, AND MICHEL RIVEILL. 2020. What do you think about your company's leaks? A survey on end-users perception towards data leakage mechanisms. *Frontiers in big Data 3*, 38.

[11] RON BITTON, KOBI BOYMGOLD, RAMI PUZIS, AND ASAF SHABTAI. 2020. Evaluating the Information Security Awareness of Smartphone Users. In *Proceedings of the 2020 CHI Conference on Human Factors in Computing Systems*, 1-13.

[12] RON BITTON, ANDREY FINKELSHTEIN, LIOR SIDI, RAMI PUZIS, LIOR ROKACH, AND ASAF SHABTAI. 2018. Taxonomy of mobile users' security awareness. *Computers & Security 73*, 266-293.

[13] Y CIFUENTES, L BELTRÁN, AND L RAMÍREZ. 2015. Analysis of Security Vulnerabilities for Mobile Health Applications. In *2015 Seventh International Conference on Mobile Computing and Networking (ICMCN 2015)*.

[14] DANIELA S CRUZES AND TORE DYBA. 2011. Recommended steps for thematic synthesis in software engineering. In *Empirical Software Engineering and Measurement (ESEM), 2011 International Symposium on* IEEE, 275-284.
[15] SERGE EGELMAN, MARIAN HARBACH, AND EYAL PEER. 2016. Behavior ever follows intention? A validation of the Security Behavior Intentions Scale (SeBIS). In *Proceedings of the 2016 CHI conference on human factors in computing systems*, 5257-5261.
[16] ADRIENNE PORTER FELT, ELIZABETH HA, SERGE EGELMAN, ARIEL HANEY, ERIKA CHIN, AND DAVID WAGNER. 2012. Android permissions: User attention, comprehension, and behavior. In *Proceedings of the eighth symposium on usable privacy and security*, 1-14.
[17] HANIA K FLATEN, CHELSEA ST CLAIRE, EMMA SCHLAGER, CORY A DUNNICK, AND ROBERT P DELLAVALLE. 2018. Growth of mobile applications in dermatology-2017 update. *Dermatology online journal 24*, 2.
[18] STEVEN FURNELL, RAWAN ESMAEL, WEINING YANG, AND NINGHUI LI. 2018. Enhancing security behaviour by supporting the user. *Computers & Security 75*, 1-9.
[19] STEVEN M FURNELL, ADILA JUSOH, AND DIMITRIS KATSABAS. 2006. The challenges of understanding and using security: A survey of end-users. *Computers & Security 25*, 1, 27-35.
[20] THOMAS GERMAIN. 2019. How to Protect Yourself From Camera and Microphone Hacking available at https://www.consumerreports.org/privacy/how-to-protect-yourself-from-camera-and-microphone-hacking-a1010757171/.
[21] VASILEIOS GKIOULOS, GAUTE WANGEN, SOKRATIS K KATSIKAS, GEORGE KAVALLIERATOS, AND PANAYIOTIS KOTZANIKOLAOU. 2017. Security awareness of the digital natives. *Information 8*, 2, 42.
[22] STEPHANIE GLEN. May 2021. Kruskal Wallis H Test: Definition, Examples & Assumptions available at https://www.statisticshowto.com/kruskal-wallis/.
[23] BRITTANY HAINZINGER. 2020. How to avoid mobile phone apps from leaking your personal data available at https://appdevelopermagazine.com/how-to-avoid-mobile-phone-apps-from-leaking-your-personal-data/.
[24] M. HUSSAIN, A. A. ZAIDAN, B. B. ZIDAN, S. IQBAL, M. M. AHMED, O. S. ALBAHRI, AND A. S. ALBAHRI. 2018. Conceptual framework for the security of mobile health applications on Android platform. *Telematics and Informatics*. DOI= http://dx.doi.org/10.1016/j.tele.2018.03.005.
[25] L. H. IWAYA, A. AHMAD, AND M. ALI BABAR. 2020. Security and Privacy for mHealth and uHealth Systems: A Systematic Mapping Study. *IEEE Access 8*, 150081-150112. DOI= http://dx.doi.org/10.1109/ACCESS.2020.3015962.
[26] K. KNORR AND D. ASPINALL. 2015. Security testing for Android mHealth apps. In *2015 IEEE 8th International Conference on Software Testing, Verification and Validation Workshops, ICSTW 2015 - Proceedings*. DOI= http://dx.doi.org/10.1109/ICSTW.2015.7107459.
[27] MURAT KOYUNCU AND TOLGA PUSATLI. 2019. Security Awareness Level of Smartphone Users: An Exploratory Case Study. *Mobile Information Systems 2019*.
[28] T. MABO, B. SWAR, AND S. AGHILI. 2018. A vulnerability study of Mhealth chronic disease management (CDM) applications (apps). In *Advances in Intelligent Systems and Computing*, 587-598. DOI= http://dx.doi.org/10.1007/978-3-319-77703-0_58.
[29] B. MARTÍNEZ-PÉREZ, I. DE LA TORRE-DÍEZ, AND M. LÓPEZ-CORONADO. 2015. Privacy and Security in Mobile Health Apps: A Review and Recommendations. *Journal of Medical Systems 39*, 1. DOI= http://dx.doi.org/10.1007/s10916-014-0181-3.
[30] FINJAN MOBILE. 2017. App Permissions – The Good, The Bad, and Why You Need to Pay Attention available at https://www.finjanmobile.com/app-permissions-the-good-the-bad-and-why-you-need-to-pay-attention/.
[31] JEFFERSON SEIDE MOLLÉRI, KAI PETERSEN, AND EMILIA MENDES. 2016. Survey guidelines in software engineering: An annotated review. In *Proceedings of the 10th ACM/IEEE international symposium on empirical software engineering and measurement*, 1-6.
[32] HEATHER MOLYNEAUX, ELIZABETH STOBERT, IRINA KONDRATOVA, AND MANON GAUDET. 2020. Security Matters… Until Something Else Matters More: Security Notifications on Different Form Factors. In *International Conference on Human-Computer Interaction* Springer, 189-205.
[33] E. P. MORERA, I. DE LA TORRE DÍEZ, B. GARCIA-ZAPIRAIN, M. LÓPEZ-CORONADO, AND J. ARAMBARRI. 2016. Security Recommendations for mHealth Apps: Elaboration of a Developer's Guide. *Journal of Medical Systems 40*, 6. DOI= http://dx.doi.org/10.1007/s10916-016-0513-6.
[34] S. MORGAN. 2020. Cybercrime To Cost The World $10.5 Trillion Annually By 2025 available at "https://cybersecurityventures.com/hackerpocalypse-cybercrime-report-2016/" [Last accessed: 06/11/2021].
[35] ALEXIOS MYLONAS, ANASTASIA KASTANIA, AND DIMITRIS GRITZALIS. 2013. Delegate the smartphone user? Security awareness in smartphone platforms. *Computers & Security 34*, 47-66.
[36] A. PAPAGEORGIOU, M. STRIGKOS, E. POLITOU, E. ALEPIS, A. SOLANAS, AND C. PATSAKIS. 2018. Security and Privacy Analysis of Mobile Health Applications: The Alarming State of Practice. *IEEE Access 6*, 9390-9403. DOI= http://dx.doi.org/10.1109/ACCESS.2018.2799522.
[37] L. PARKER, V. HALTER, T. KARLIYCHUK, AND Q. GRUNDY. 2019. How private is your mental health app data? An empirical study of mental health app privacy policies and practices. *International Journal of Law and Psychiatry 64*, 198-204. DOI= http://dx.doi.org/10.1016/j.ijlp.2019.04.002.
[38] KATHRYN PARSONS, DRAGANA CALIC, MALCOLM PATTINSON, MARCUS BUTAVICIUS, AGATA MCCORMAC, AND TARA ZWAANS. 2017. The human aspects of information security questionnaire (HAIS-Q): two further validation studies. *Computers & Security 66*, 40-51.

[39] WEI PENG, SHAHEEN KANTHAWALA, SHUPEI YUAN, AND SYED ALI HUSSAIN. 2016. A qualitative study of user perceptions of mobile health apps. *BMC Public Health 16*, 1, 1158.
[40] M. PLACHKINOVA, S. ANDRES, AND S. CHATTERJEE. 2015. A Taxonomy of mHealth apps - Security and privacy concerns. In *Proceedings of the Annual Hawaii International Conference on System Sciences*, 3187-3196. DOI= http://dx.doi.org/10.1109/HICSS.2015.385.
[41] LINDSAY RAMEY, CANDICE OSBORNE, DONALD KASITINON, AND SHANNON JUENGST. 2019. Apps and Mobile Health Technology in Rehabilitation: The Good, the Bad, and the Unknown. *Physical Medicine and Rehabilitation Clinics 30*, 2, 485-497.
[42] ABBAS RAZAGHPANAH, RISHAB NITHYANAND, NARSEO VALLINA-RODRIGUEZ, SRIKANTH SUNDARESAN, MARK ALLMAN, CHRISTIAN KREIBICH, AND PHILLIPA GILL. 2018. Apps, trackers, privacy, and regulators: A global study of the mobile tracking ecosystem. In *The 25th Annual Network and Distributed System Security Symposium (NDSS 2018)*.
[43] ELISSA M REDMILES, YASEMIN ACAR, SASCHA FAHL, AND MICHELLE L MAZUREK. 2017. *A summary of survey methodology best practices for security and privacy researchers.*
[44] ALLIED MARKET RESEARCH. 2021. Digital Health Market available at https://www.alliedmarketresearch.com/digital-health-market-A10934.
[45] LAERD STATISTICS. 2018. Chi-Square Test for Association using SPSS Statistics [online] available at https://statistics.laerd.com/spss-tutorials/chi-square-test-for-association-using-spss-statistics.php.
[46] LAERD STATISTICS. 2018. Kruskal-Wallis H Test using SPSS Statistics [online] available at https://statistics.laerd.com/spss-tutorials/kruskal-wallis-h-test-using-spss-statistics.php.
[47] ERIC STRUSE, JULIAN SEIFERT, SEBASTIAN ÜLLENBECK, ENRICO RUKZIO, AND CHRISTOPHER WOLF. 2012. PermissionWatcher: Creating User Awareness of Application Permissions in Mobile Systems. In *International Joint Conference on Ambient Intelligence* Springer, 65-80.
[48] KEITH S TABER. 2018. The use of Cronbach's alpha when developing and reporting research instruments in science education. *Research in science education 48*, 6, 1273-1296.
[49] MOHSEN TAVAKOL AND REG DENNICK. 2011. Making sense of Cronbach's alpha. *International journal of medical education 2*, 53.
[50] TERMLY LEGAL TEAM. 2017. Mobile App Privacy Policy Template available at "https://termly.io/resources/templates/app-privacy-policy/#what-is-an-app-privacy-policy".
[51] IPC TECHNOLOGIES. 2021. Hackers increasingly exploit human factor avaliable at https://www.ipctech.com/hackers-increasingly-exploit-human-factor/.
[52] G. THAMILARASU AND C. LAKIN. 2017. A security framework for mobile health applications. In *Proceedings - 2017 5th International Conference on Future Internet of Things and Cloud Workshops, W-FiCloud 2017*, 221-226. DOI= http://dx.doi.org/10.1109/FiCloudW.2017.96.
[53] ROGER TOURANGEAU AND TOM W SMITH. 1996. Asking sensitive questions: The impact of data collection mode, question format, and question context. *Public opinion quarterly 60*, 2, 275-304.
[54] ROGER TOURANGEAU AND TING YAN. 2007. Sensitive questions in surveys. *Psychological bulletin 133*, 5, 859.
[55] UCLA. 2021. What does Cronbach's alpha mean? available at "https://stats.idre.ucla.edu/spss/faq/what-does-cronbachs-alpha-mean/".
[56] RICK WASH, EMILEE RADER, AND CHRIS FENNELL. 2017. Can people self-report security accurately? Agreement between self-report and behavioral measures. In *Proceedings of the 2017 CHI conference on human factors in computing systems*, 2228-2232.
[57] BRYAN WATSON AND JUN ZHENG. 2017. On the user awareness of mobile security recommendations. In *Proceedings of the SouthEast Conference*, 120-127.
[58] PRIMAL WIJESEKERA, ARJUN BAOKAR, ASHKAN HOSSEINI, SERGE EGELMAN, DAVID WAGNER, AND KONSTANTIN BEZNOSOV. 2015. Android permissions remystified: A field study on contextual integrity. In *24th {USENIX} Security Symposium ({USENIX} Security 15)*, 499-514.
[59] FATIMA ZAHRA, AZHAM HUSSAIN, AND HASLINA MOHD. 2018. Factor Affecting Mobile Health Application for Chronic Diseases. *Journal of Telecommunication, Electronic and Computer Engineering (JTEC) 10*, 1-11, 77-81.
[60] MINE ZEYBEK, ERCAN NURCAN YILMAZ, AND İBRAHIM ALPER DOĞRU. 2019. A Study on Security Awareness in Mobile Devices. In *2019 1st International Informatics and Software Engineering Conference (UBMYK)* IEEE, 1-6.
[61] L. ZHOU, J. BAO, V. WATZLAF, AND B. PARMANTO. 2019. Barriers to and facilitators of the use of mobile health apps from a security perspective: Mixed-methods study. *JMIR mHealth and uHealth 7*, 4. DOI= http://dx.doi.org/10.2196/11223.
[62] F. ZUBAYDI, A. SALEH, F. ALOUL, AND A. SAGAHYROON. 2015. Security of mobile health (mHealth) systems. In *2015 IEEE 15th International Conference on Bioinformatics and Bioengineering, BIBE 2015*. DOI= http://dx.doi.org/10.1109/BIBE.2015.7367689.